\documentclass{PoS}

\usepackage{amsmath}
\usepackage{wrapfig}
\title{Mass Splittings in a Linear Sigma Model}

\ShortTitle{Mass Splittings}

\author{Diego de Floor\\
        University of Iowa Department of Physics and Astronomy\\
        E-mail: \email{diego-defloorsilva@uiowa.edu}}

\author{\speaker{Erik Gustafson}\\
        University of Iowa Department of Physics and Astronomy\\
        E-mail: \email{erik-j-gustafson@uiowa.edu}}
        
\author{Yannick Meurice\\
        University of Iowa Department of Physics and Astronomy\\
        E-mail: \email{yannick-meurice@uiowa.edu}}

\abstract{
	We derived the tree level spectrum to an extension to the linear sigma model, proposed by \cite{Meurice:2017zng}, describing an EFT for an $SU(3)_c$ gauge theory with $N_f$ flavors of fermions and $N_1$ fermions have a mass $m_l$ and $N_2$ fermions have a mass $m_h$. We examined the effects of a small mass splitting on single mass data for 8 and 12 flavors of fermions corresponding to the unperturbed case. Our intention is to encourage more simulations of split mass theories with 8, 10 and 12 flavors of fermions.}

\FullConference{The 36th Annual International Symposium on Lattice Field Theory - LATTICE2018\\
		22-28 July, 2018\\
		Michigan State University, East Lansing, Michigan, USA.}

\begin{document}

\section{Introduction}
Recent lattice calculations [2-6] have found a light $\sigma$ in SU(3) gauge theories with 8 and 12 degenerate flavors. Current studies of the $\eta'$ support the possibility that the breaking of the $U(1)_A$ symmetry, which depends in a distinct way on $N_f$ can explain the $\sigma$ becoming lighter as $N_f$ increases [1]. It would be interesting to investigate whether the unexpectedly light states (e.g. $\sigma$) persist when some of the fermions have a greater mass. Part of the spectrum for the  4+8 spectrum has been extracted numerically [6,7]. 
\section{Lagrangian}
We begin by defining a set of fields $\phi_{ij}$, as in \cite{Meurice:2017zng}, which are $N_f\times N_f$ matrices transforming as  $\bar{\psi}_{Rj}\psi_{Li}$ under $U(N_f)_L \otimes U(N_f)_R$. We use the parameterization for the fields 
\begin{equation}
\phi=(S_\alpha+iP_\alpha)\Gamma^\alpha,
\end{equation}
where the sum over $\alpha = 0, 1, ... , N_f^2 - 1$ is for a basis of $N_f \times N_f$ Hermitian matrices $\Gamma^{\alpha}$ such that: $ Tr(\Gamma^{\alpha}\Gamma^{\beta}) = (1/2)\delta^{\alpha \beta}$ and define $\Gamma_0=1/\sqrt{2N_f}$ 
\large
$\mathbf{1}$
%$\mathbb{1}$ 
\normalsize
$_{N_f\times N_f}$. The Lagrangian for this system is defined as 
\begin{equation}
\mathcal{L} = Tr(\partial_{\mu} \phi \partial^{\mu} \phi^{\dagger}) + V_0 + V_a + V_m,
\end{equation}
where the first term is a canonical kinetic term and the remaining three components correspond to various breakings of the symmetries. The $V_0$ term corresponds to a $U(N_f)_L \otimes U(N_f)_R$ invariant expression:
\begin{equation}
V_0 = - \mu^2 Tr(\phi \phi^{\dagger})+ \frac{\lambda_{\sigma} - \lambda_{a0}}{2}(Tr(\phi \phi^{\dagger}))^2 + \frac{\lambda_{a0} N_f}{2} Tr((\phi \phi^{\dagger})^2).
\end{equation} 
The $V_a$ term corresponds to the breaking of the axial $U(1)_A$ symmetry and has the form:
\begin{equation}
V_a = - 2 (2 N_f)^{N_f/2 - 2} X (det(\phi) + det(\phi^{\dagger})).
\end{equation}
The $V_m$ term, which is defined,
\begin{equation}
V_m = - Tr(\mathcal{M} (\phi + \phi^{\dagger})),
\end{equation} corresponds to a breaking of $SU(N_f)_V$ symmetry into a $SU(N_1)_v\otimes SU(N_2)_v$ and where,
\begin{equation}
\mathcal{M} = b_0 \Gamma^0 + b_8 \Gamma^8
\end{equation} 

\section{Spectrum}
The mass spectrum for this theory arises from the second derivatives of the potential with respect to the scalar and pseudoscalar fields. The pseudoscalar non-singlet spectrum can be compactly written as:
\begin{equation}
\begin{split}
M_{\pi_{ll}}^2 &= -\mu^2 + \frac{\lambda_{\sigma}-
\lambda_{a0}}{2}(N_1 v_1^2 + N_2 v_2^2) +\lambda_{a0}/2v_1^2 - X/N_f v_1^{N_1 - 2} v_2^{N_2}\\
M_{\pi_{lh}}^2 -M_{\pi_{ll}}^2 &= \Big(\lambda_{a0}/2 +  X/N_f v_1^{N_1-2}v_2^{N_2 - 2}\Big)(v_2-v_1)v_2\\
M_{\pi_{hh}}^2 - M_{\pi_{ll}}^2 &= \Big(\lambda_{a0}/2 +  X/N_f v_1^{N_1 - 2}v_2^{N_2 - 2}\Big)(v_2^2 - v_1^2)
\end{split}
\end{equation}
and the scalar spectrum spectrum can be written as: 
\begin{equation}
\begin{split}
M^2_{a0_{ll}} - M_{\pi_{ll}}^2& = \lambda_{a0} v_1^2 + 2 X/N_f v_1^{N_1 - 2} v_2^{N_2}\\
M^2_{a0_{lh}} - M_{\pi_{lh}}^2 &= \lambda_{a0} v_1 v_2+ 2 X/N_f v_1^{N_1 - 1} v_2^{N_2-1}\\
M^2_{a0_{hh}} - M_{\pi_{hh}}^2 &= \lambda_{a0} v_2^2 + 2 X/N_f v_1^{N_1} v_2^{N_2-2}
\end{split}
\end{equation}
where $v_1$ and $v_2$ are related to the vaccuum expectation values of $S_0$ and $S_8$  by the relations: $\langle S_0 \rangle = (1/N_f)(N_1v_1+N_2v_2)$ and $ 
\langle S_8 \rangle = (\sqrt{N_1N_2}/N_f)(v_1-v_2)$.
\section{Unperturbed data}
\begin{figure}
\includegraphics[scale=0.45]{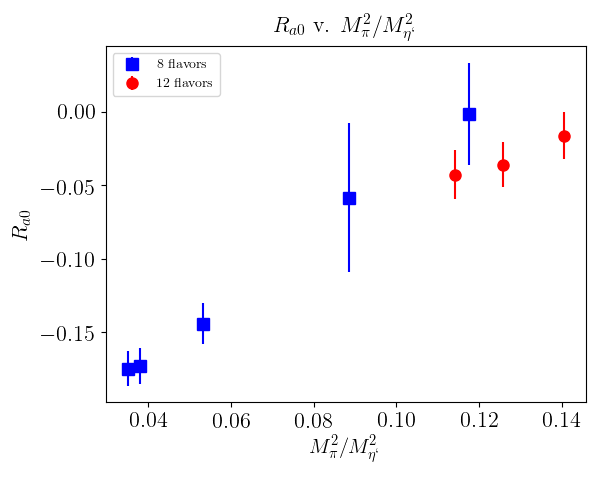}
\includegraphics[scale=0.45]{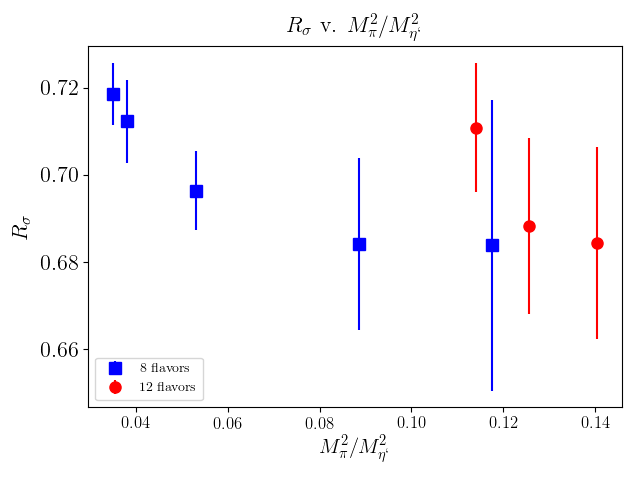}
\caption{Numeric values for coupling rations $R_{\sigma}$ and $R_{a0}$ for $N_f = 8,12$ using data from \cite{Aoki:2014oha,Aoki:2016wnc,Aoki:2013zsa}}
\label{fig:couplings}
\end{figure}

Before examining how the spectrum splits changes by small mass splittings, we want to examine both how the couplings change with respect to the mass of the fermions, and how the masses change as the chiral limit of the theory is approached. The couplings ($mu$,$\lambda_{\sigma}$,$\lambda_{a0}$ etc.) are likely non-trivially related to the fermion mass and so it is important to examine how drastic the changes in these couplings are. In terms of the unperturbed spectrum derived in \cite{Meurice:2017zng} it is possible to write the couplings as ratios of the spectroscopic data:
\begin{equation}
\begin{split}
R_{a0} &= \lambda_{a0} v^2 / M_{\eta'}^2 = ((M_{a0}^2 - M_{\pi}^2) - (2/N_f) (M_{\eta'}^2 - M_{\pi}^2))/M_{\eta'}^2\\
R_{\sigma} &= \lambda_{\sigma} v^2 / M_{\eta'}^2 = ((M_{a0}^2 - M_{\pi}^2) +(1 - (2/N_f)) (M_{\eta'}^2 - M_{\pi}^2))/M_{\eta'}^2.
\end{split}
\end{equation}
We can see in figure \ref{fig:couplings} that the couplings vary slowly over the range of fermion masses that are investigated. 

\begin{figure}
\includegraphics[scale=0.45]{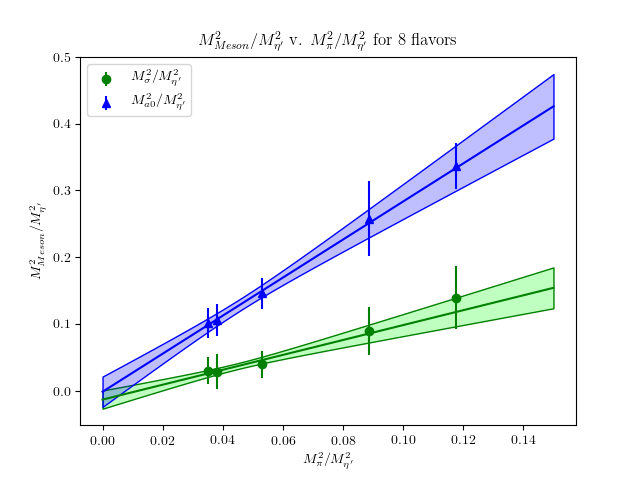}
\includegraphics[scale=0.45]{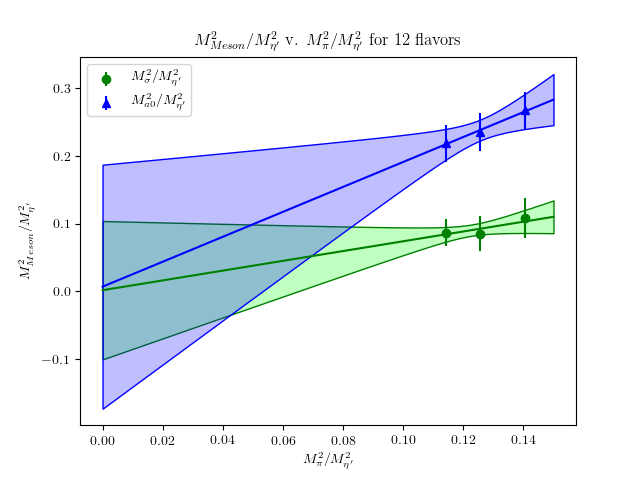}
\caption{chiral extrapolation of the meson masses for 8 (left) and 12 (right) flavors }
\label{fig:chiral}
\end{figure}

When we extrapolate the spectrum toward the chiral limit (figure \ref{fig:chiral}), we find a surprising result that the $\sigma$ and $a0$ masses for both the 8 and 12 flavor cases extrapolate close to zero or less than zero. Although the uncertainty in the 12 flavor case is significant, the uncertainties in the 8 flavor case are quite small and this suggests that either the model breaks down near the chiral limit or some unexpected behaviors namely a small a0 mass would be present and this would go contrary to our expectations that the heavier states such as the $a0$ could be integrated out. 

\section{Perturbative Results}

\begin{figure}
\includegraphics[scale=0.9]{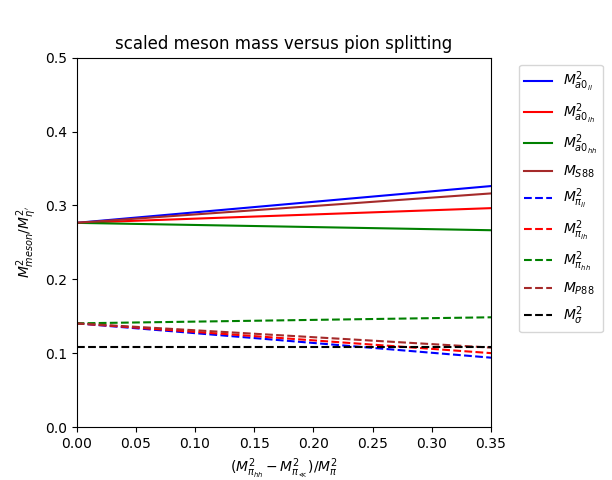}
\caption{linear approximation of the mass splittings for one choice of fermion mass using $amf=0.05$ and $N_1 = 2$ and $N_2 = 10$}
\label{fig:masses}
\end{figure}

We can linearly approximate the mass-split spectrum using the unperturbed masses by defining the seperation: $\Delta_{meson} = M_{meson}^2 - M_{\pi}^2$. Using this definition we can write the splittings for the pseudoscalars:
\begin{table}
\centering
\begin{tabular}{|c|c|c|c|}
\hline
\hline
$N_f$ & $(a m_f)$ &  $\frac{\Delta_{a_0}}{M_{\eta'}^2}$ & $\frac{3\Delta_{a_0} - (8/N_f)\Delta_{\eta'}}{M_{\eta'}^2}$\\\hline
8 & 0.012 & 0.0584(74) &  -0.669(71) \\\hline 
8 & 0.015 & 0.0644(78) &  -0.724(81) \\\hline 
8 & 0.02 & 0.0885(95) &  -0.640(70) \\\hline 
8 & 0.03 &  0.160(41) & -0.38(16) \\\hline 
8 & 0.04 &  0.214(28) &  -0.22(11) \\\hline 
12 & 0.04 &  0.0854(94) & -0.225(52) \\\hline 
12 & 0.05 & 0.1079(65) & -0.163(51) \\\hline 
12 & 0.06 & 0.1124(73)& -0.261(65) \\\hline 
%\caption{Values of}% $\frac{\Delta_{a_0}}{M_{\eta'}^2}$}% and $\frac{3\Delta_{a_0} - (8/N_f)\Delta_{\eta'}}{M_{\eta'}^2}}$.}
\end{tabular}
\caption{Values  $\frac{\Delta_{a_0}}{M_{\eta'}^2}$ and $\frac{3\Delta_{a_0} - (8/N_f)\Delta_{\eta'}}{M_{\eta'}^2}$ using the LatKMI data [3]}
\label{tab:del}
\end{table}
\begin{equation}
\begin{split}
M_{\pi_{hh}}^2 - M_{\pi_{ll}}^2 &\simeq \frac{v_2 - v_1}{v} \Delta_{a0}\\
M_{\pi_{lh}}^2 - M_{\pi_{ll}}^2 &\simeq \frac{1}{2}(M_{\pi_{hh}}^2 - M_{\pi_{ll}}^2)\\
M_{P_{88}}^2 - M_{\pi_{ll}}^2 &\simeq \frac{N_1}{N_f}(M_{\pi_{hh}}^2 - M_{\pi_{ll}}^2),
\end{split}
\end{equation}
and the scalars:
\begin{equation}
\begin{split}
M_{a0_{hh}}^2 - M_{a0_{ll}}^2 &\simeq \frac{v_2 - v_1}{v}(3 \Delta_{a0} - \frac{8}{N_f}\Delta_{\eta'})\\
M_{a0_{hl}}^2 - M_{a0_{ll}}^2 &\simeq \frac{1}{2}(M_{a0_{hh}}^2 - M_{a0_{ll}}^2)\\
M_{S_{88}}^2 - M_{a0_{ll}}^2 &\simeq \frac{N_1}{N_f}(M_{a0_{hh}}^2 - M_{a0_{ll}}^2).
\end{split}
\end{equation}
The numeric values for the splittings are listed in table \ref{tab:del}. One of the suprising results is that a0's undergo an inversion ($M_{a0_{ll}}^2 > M_{a0_{lh}}^2 > M_{a0_{hh}}^2$), which is contrary to what is expected from QCD. A plot of the linear approximations are shown in \ref{fig:masses}; we have only used a linear approximation because the couplings, $\lambda_{a0}$, $\lambda_{\sigma}$ and X all evolve as the fermion mass changes. In order to examine any effects beyond first order, these non-linear effects to the couplings must be taken into account.

\section{Conclusions}
 For a mass split linear sigma model we have developed a perturbative expansion in the mass difference $(M_{\pi_{hh}}^2 - M_{\pi_{ll}}^2)$ which provides simple results for differences of squared masses. We encounter a surprising result when the familiar ordering for pseudoscalars ( $M_{\pi_{hh}}^2 > ... >M_{\pi_{ll}}^2$) is imposed the ordering of the scalars is reversed ( $M_{a0_{ll}}^2 > M_{a0_{hl}}^2 > M_{a0_{hh}}^2$). We suggest examining this inversion for $N_1 = 2$ and $N_2 = 6$ and $am_1 = 0.012$ and $am_2 = 0.015$ which are parameters used by the LatKMI collaborations. The masses are small enough to have a clearly negative $\lambda_{a0}$ and relative mass difference small enough to avoid large non-linear corrections and would be an effective first step in examining whether or not this inversion actually occurs.

Acknowledgements: We would like to thank O. Witzel and A. Gasbarro for discussions. This research was supported in part by Department of Energy under Award Numbers DOE grant DE-SC0010113

\end{document}